\documentclass[twoside]{article}
\usepackage{acronym} 
\usepackage[sc]{mathpazo} 
\usepackage[T1]{fontenc} 
\usepackage{microtype} 
\usepackage[hmarginratio=1:1,top=32mm,columnsep=20pt]{geometry} 
\usepackage{multicol} 
\usepackage{hyperref} 
\usepackage[hang, small,labelfont=bf,up,textfont=it,up]{caption} 
\usepackage{booktabs} 
\usepackage{float} 
\usepackage{lettrine} 
\usepackage{paralist} 
\usepackage{abstract} 
\usepackage{titlesec} 
\titleformat{\section}[block]{\large\scshape\centering}{\thesection.}{1em}{} 
\usepackage{fancyhdr} 
\usepackage[T1]{fontenc}
\usepackage[utf8]{inputenc}
\usepackage{authblk}
\usepackage{graphicx}
\usepackage{hyperref}
\usepackage{color}
\usepackage{graphicx}
\usepackage{dcolumn}
\usepackage{color}
\usepackage{enumerate}
\usepackage{multirow}
\begin{document}

\title{A Framework for Reproducible, Interactive Research: Application to health and social sciences} 

\author[1]{Joao Ricardo Nickenig Vissoci}
\author[2,3]{Clarissa G. Rodrigues}
\author[4]{Luciano de Andrade}
\author[5]{Jose Eduardo Santana}
\author[6]{Amrapali Zaveri}
\author[2]{Ricardo Pietrobon}

\affil[1]{Inga College - Medicine Department}
\affil[2]{Duke University Medical Center, \{jnv4, clarisa.rodrigues, rpietro\}@duke.edu}
\affil[3]{Instituto de Cardiologia do RS, Fundacao Universitaria de Cardiologia, Rio Grande do Sul, Brazil}
\affil[4]{Universidade Estadual do Oeste, do Parana, Brazil, luciano.andrader@unioeste.br}
\affil[5]{Instituto de Computação, Universidade Federal de Alagoas (UFAL), Alagoas, Brazil, jes@ic.ufal.br}
\affil[6]{Universit\"at Leipzig, Institut f\"ur Informatik,
D-04103, Leipzig, Germany, zaveri@informatik.uni-leipzig.de}

\maketitle 
\thispagestyle{fancy} 

\begin{abstract}
The aim of this article is to introduce a reporting framework for reproducible, interactive research applied to Big Clinical Data, based on open source technologies. The framework is constituted by the following three axes: (i) data, (ii) analytical codes and (iii) dissemination. In this paper, different documentation formats and online repositories are introduced. To integrate and manage the reproducible contents, we propose the R Language as the tool of choice. All the information is then published and gathered in a website for different projects. This framework is free and user friendly and is proposed to enhance reproducibility of health-science reports.
\end{abstract}


\begin{multicols}{2}

\section{Introduction}
\lettrine[nindent=0em,lines=3]{W} ith the growing amount of data in healthcare, the ability to analyze large datasets and report results adequately has become a key factor of research and innovation~\cite{Manyika}, which supports the creation of new technologies and improved clinical decision making. The increased complexity of these datasets brings together difficulties and new challenges in terms of data management, modeling and communication. Therefore, investigators are now focusing on developing reproducible research protocols including entirely reproducible data analysis. It implies that the results reported in a publication can be immediately reproduced by granting access to both the datasets as well as the statistical and data mining scripts of the study~\cite{Peng}  

In order to make the information widely usable, the value of data collection, analysis and communication as well as the use of common standards for sharing information have been recognized. In addition to increasing dissemination and better understanding of research findings, data sharing can also support confirmation or refutation of research by allowing replication and increased transparency of results~\cite{Peng,Groves} 

However, data sharing does bring some implementation challenges and possible risks. Potential invasion of participants privacy and breaking of patients confidentiality are primary concerns when making datasets public. Secondly, adequate data management, academic and commercial primacy, and intellectual property rights as well as journal copyrights are factors to be careful with while publishing data~\cite{Groves}. In this context, the use of an adequate framework becomes essential to allowing reproducible research without compromising such aspects, specially when analyzing and reporting results from large datasets.

Thus, the aim of this article is to introduce a simple reporting framework for \textit{reproducible, interactive research} applied to health and social scienc.
The framework is constituted by the following three axes: (i) data (Section~\ref{data}), (ii) analytical codes (Section~\ref{scripts}) and (iii) dissemination (Section~\ref{dissemination}). In this paper, different documentation formats and online repositories are introduced. To integrate and manage the reproducible contents, we propose the R Language as the tool of choice. All the information is then published and gathered in a website for different projects. This framework is free and user friendly and is proposed to enhance reproducibility of health-science reports.

\section{Reproducible Research Framework}
The framework proposed in this paper is based on the concept that an appropriate reproducible research report should allow one to totally reproduce the methods applied. Thus, we understand that besides making the analytical data, code and figures available, an adequate reproducible research framework should integrate tools and features in a way that others could reach the same results and understand the process behind it.

Therefore, in order to achieve an adequate integration between data, codes and outcomes (figures, tables, numerical results and others) in our framework, we utilize the R Language~\cite{R} as the central tool. R has the ability of integrating and managing different data formats, codes and formats. In addition, it allows communication with several other analytical softwares such as SAS, Stata and SPSS~\cite{IBM, SAS, Stata}.
 
\subsection{Data formats}
\label{data}
The first issue about making a research protocol reproducible is the data management process. There are several ways of storing data and many different data formats. In our perspective, some of them are better by allowing integration with data analysis softwares and online repositories as well as their ease of use. In the following sections, we demonstrate some of these formats we have been using and their integration with our reproducible Framework.

\subsubsection{Reproducible Data}
When making datasets publicly available, one must be concerned with the information that is going to be made public. In this context, the Health Insurance Portability and Accountability Act (HIPAA) developed a section on Protected Health Information (PHI), which means that individually identifiable health information must be kept confidential when sharing data in healthcare. The complete list of PHI can be found at the Health and Human Services US Department~\cite{HIPPA}

Secondly, it is important to make sure that the data is coded with appropriate names that allows other people to read and understand the content easily. To make it easier, we strongly encourage the publication of a complete and organized \textit{data dictionary} together with the dataset, containing variable labels, respective code, data characteristics (continuous, discrete, ordinal, dichotomous, etc.) and any other source of relevant information (e.g. length of Likert scale, categorization factors).

\subsubsection{CSV} 
Comma separated values (CSV) is a format readily available for consumption by any data analysis language or software. 
However, it does not provide a way to update the data once it is downloaded other than downloading the dataset again. In addition, the CSV format does not offer any security features. 

On the other hand, CSV files have one of the best usability  experiences among all the formats and it can be easily integrated with R using online repositories (i.e. Google Drive or Dryad) through different R packages. One such package is the RCurl package~\cite{Lang}, which can integrate R with different HTML domains, among them a .csv spreadsheet from Google Docs.
    
\subsubsection{RDF, LOD and SPARQL endpoints}
Semantic Web technologies have recently become popular given the success provided by Linked Open Data (LOD)~\cite{footnote}. 
The data is represented with the help of the Resource Description Framework (RDF) format, while data sets themselves are queried through the SPARQL (a recursive acronym for SPARQL Protocol and RDF Query Language).  
Main advantages include the data availability 24/7 with automated updates and also the ability to dynamically merge across data sets sharing identical elements (classes or instances). 

RDF data can be easily integrated with R Analytical codes through the RRDF package~\cite{Williighagen}. This package allows users to perform SPARQL queries inside R's workspace. In addition to this package mentioned there is a whole set of tutorials and packages that can be used within R~\cite{R}.

\subsubsection{JSON}
JavaScript Object Notation (JSON) is considered on of the best data-interchange formats.
It is a text format with conventions familiar to several programming languages such as C++, Java, JavaScript and Python. 

More information and specifications about how to integrate JSON data with specific applications can be found at~\cite{json}. It's connection with R analytical code is executed through the rjson package \cite{Rproject json} which converts JSON objects into R objects.

\subsection{Data repositories}
After deciding a format of data to be used, it is also mandatory to use an online repository to store the data and integrate it with the analytical codes (discussed later in Section~\ref{scripts}). In the following sections, we present some options of free repositories that are used by our group.

\subsubsection{Dryad}
Dryad~\cite{Dryad} is an international repository specific for data related to scientific publications. It allows data to be deposited easily and readily provides the citation related to the respective publication. Dryad can be integrated with R, thus improving interoperability~\cite{Chamberlain}. 

\subsubsection{Figshare}
Figshare~\cite{Figshare}  is an online repository, similar to Dryad, that allows researchers to choose a publication with the ability to be cited within the paper.

Additionally, Figshare supports not only data but also other types of research outputs such as figures, datasets, media files, papers, posters or even file sets with different types of documents. A major advantage of Figshare is its ability of easily sharing and discovering information about different research projects. We have used Figshare to publish datasets (in .csv formats) as well as figures. Examples can be found in~\cite{Moreira,DalPonte}.
In addition, Figshare can also be integrated with R through some packages~\cite{rfigshare}. 

\subsubsection{Google drive}
Google Drive is another online repository which facilitates collaboration and sharing of files~\cite{Google}. This application from Google integrates texts, spreadsheets, presentations and other editors from Google (i.e. Google Docs, Google Sheets, Google Forms and others) and also allows the user to store forms, drawing, and different types of files in the cloud.

Google Drive is extensively used to share data, codes and other outputs among researchers in our group~\cite{ROR}. One example to connect data stored in Google Drive with R is the RCurl package . This package allows users to compose general HTTP requests and call URLs and other web formats, such as datasets in .csv format. Another way is to simply open the files stored in Google Drive (Spreadsheets or R-Scripts, for example) inside R, through RStudio~\cite{rstudio}.

In addition, we also use Google Drive as a way to integrate and facilitate collaborative writing and coding in R, since this approach has been found more user friendly to content researchers than other more sofisticated repositories.

\subsection{Analytical scripts}
\label{scripts}
Publishing analytical codes is an important step in a reproducible framework besides the connection between the codes and the data. Therefore, we demonstrate here the different software that can be used to generate, publish and manage the analytical codes.

\subsubsection{R Language Statistical Software}
As mentioned before, R~\cite{R}  is the central tool of our reproducible research framework. As a definition, R is an open source software for statistical analysis and graphic creation. It has been developed by a vast community of collaborators from several countries and institutions.

Although R is not superior to other statistical softwares in every aspects (such as intuitive GUI interface, or pre-defined operations), it gathers qualities which makes it a better option to our framework than other statistical environments. One major advantage of R is its collaborative function in the development of packages. R has a huge library~\cite{R} {(Comprehensive R Archive Network - CRAN)} of packages for statistical analysis, graphic creation, data mining and management, and integration with other softwares and programming languages.  

This collaborative ability, besides making R a powerful analytical environment, makes it assume a position in our framework as a glue for other languages and technologies such as Python, Java, relational databases, RDF, C, C++, Weka, among many others. This way we can gather data and data storage tools, analytical coding and repositories for outputs, making a research project fully reproducible. In addittion, R has being used by a large community, and has a lot of references to lean on.

In our group we opt to run R through RStudio~\cite{rstudio}. This platform is also open source and is an integrated environment that helps to visualize the different R interfaces (workspace, graph, scripts and log). Other than that, it facilitates the management of multiple working directories through the definitions of projects.

\subsubsection{Reproducible Scripts}
As suggested by Hadley in his github repository~\cite{hadley}, the idea is to create a code that can be recreated just by copying the codes we publish online. Therefore, each code must be connected to the dataset and contain all the information needed to be performed.

The elements of a reproducible script in R include the required packages, connection to the data, codes and codes descriptions. Each function in R is called upon a package where it is nested. So, for anyone else to be able to reproduce our codes, she must have all the packages installed.

Regarding the data, we have already discussed earlier the possible formats and ways to publish it. It is noteworthy that the data must be aligned with the codes. This means that all the variables must be named exactly with the names used in the codes. Also, every data management information must be inserted in the codes so that whoever is trying to reproduce it might reach the same results. Finally, each line must have a description of its purpose and use. 

\subsubsection{Github} 
Github~\cite{github} is an online repository built to facilitate the collaborative writing of computing codes. It not only allows the sharing of codes  but facilitates collaboration through the copy (hereby called \textit{"fork"}) of project pages in a safer way, regarding the original code. Among all the qualities of using Github as a reproducible strategy in the analytical coding process, we highlight its strong connectivity with R. It allows not only the sharing and management of codes in websites, but also simulates R outputs with kntir~\cite{knitr}.
 
There are several possibilities of using R integrated with Github. We have been using Github mainly to: 
\begin{itemize}
\item publish analytical R-Scripts
\item promote collaboration among our data analysts when creating or debugging data analysis
\item generate automatic data reports for open design projects
\item create templates for data analysis (hereby called data analysis toolbox) with explanation of the methods (using wiki pages) and description of codes and outcomes.
\end{itemize}

\subsection{Dynamic research}
In order to have a complete reproducible script and also to facilitate data dissemination and visualization, it is important to obtain automated and dynamic representation of tables, figures and reports. R allows the creation of analytical codes that generates automated reports, such as the knitr package~\cite{knitr}, which translates the analysis into an HTML report (or other formats such a PDF). In summary, this package translates the code into a report mixing Latex and markdown languages. An example of its application can be found in our Github repository~\cite{rpietro} for the, Glocal Open Design Collection project. In this specific project we used knitr associated with a R code to generate an automated report about data quality and associations.

Another way of using R to generate dynamic research is by developing interactive graphs. These are graphs that might be customized or modified by the user (research subject, patient or any other stakeholder) to get different slices of the dataset. R has several ways of generating interactive graphs. Here, we would like to introduce rggobi and Shiny~\cite{Lang, Adler, Shiny}. However, there are options that can be found at the CRAN task view for dynamic graphs~\cite{Koh} 

\subsection{Licensing}
Since all the documentation we are using is going to be made public we need to assure that its use is covered by a license. This will assure that any use other than that allowed by the license, is not performed by the users. This is fairly important due to the relevance of the information being made public. In our framework, we have used Creative Commons, which is a free copyright license framework~\cite{creative}.
  
Inserting a line regarding the licensing characteristics in each of the documents in a project is sufficient to specify the type of license. The licensing assures the need for approval from the copyright owner. Basically, we allow the user to share and adapt the specific parts of the project. The only restriction is that the user must attribute the documents to the original authors and must use it only for noncommercial purposes.

Examples of licenses are: This code is licensed under a Creative Commons Attribution - Noncommercial 3.0 Unported License. You are free: to Share - to copy, distribute and transmit the work, to Remix - to adapt the work, under the following conditions: Attribution - You must attribute the work in the manner specified by the author or licensor (but not in any way that suggests that they endorse you or your use of the work). Noncommercial - You may not use this work for commercial purposes. With the understanding that: Waiver - Any of the above conditions can be waived if you get permission from the copyright holder. Public Domain - Where the work or any of its elements is in the public domain under applicable law, that status is in no way affected by the license. Other Rights - In no way are any of the following rights affected by the license: Your fair dealing or fair use rights, or other applicable copyright exceptions and limitations; The author's moral rights; Rights other persons may have either in the work itself or in how the work is used, such as publicity or privacy rights. Notice - For any reuse or distribution, you must make clear to others the license terms of this work. The best way to do this is with a link to this web page. For more details see \url{http://creativecommons.org/licenses/by-nc/3.0/}~\cite{creative}

\subsection{Data Dissemination and Communication}
\label{dissemination}
Other than discussing methods and tools to make a research reproducible, we also believe that it is important to include a facilitation of the data communication and dissemination. This will not only allow users to access the research project but will also catalyze the reach and dissemination of the respective projects. 

In order to disclose and gather all the material from our groups' research projects that was made public, we created websites (using Google Sites~\cite{Googlesites}) for each of the projects where we included links to data repositories, code repositories and inserted reports and graphs. Any web design tool can be used but our choice for Google Sites is based on its free access and user friendly interface. An example is the Observer Agreement website which integrates all the reproducible documentation for our researchers with observer agreement about orthopedic scales projects~\cite{observeragreement}. 

\subsection{Overall workflow}
Summaryzing the information discussed, we created a simple graphical demonstration of the framework's conception (Figure 1). As mentioned before, R Languge software gets a highlighted position in the framework's model. So R is used to manage and coordinate the documentation. Data is stored in open access online repositories,in a R supported formats that will allow the connection between data and analytical code. The analytical codes are developed within R interface and stored in a open access online repositiry. Outputs generated by the codes are also stored in open sourced online repositores. All this information is licensed and integrated in a website for the research project.

\begin{figure*}[htb]
\centering
\includegraphics[scale=0.5]{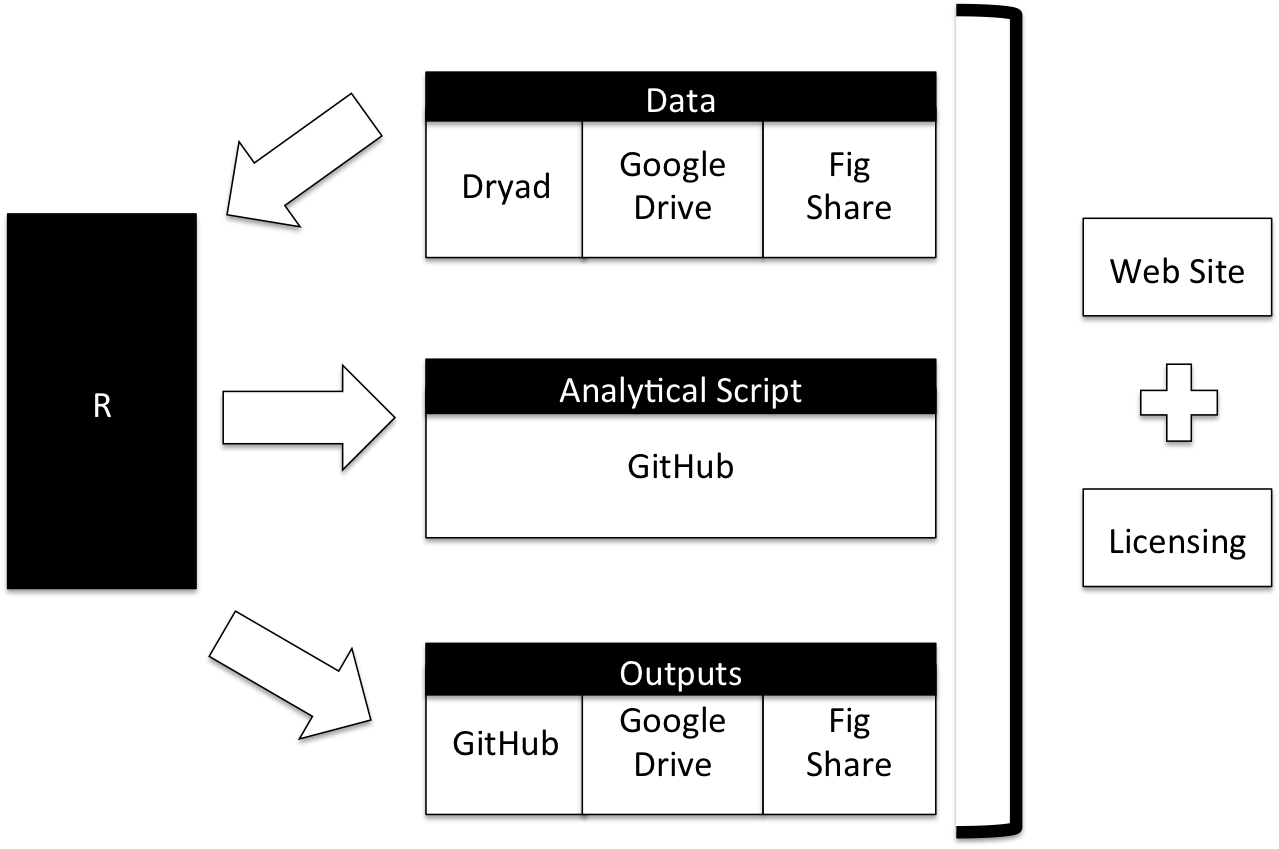}
\caption{Depiction of the reproducible research framework.}
\label{fig:workflow}	
\end{figure*}

\section{Discussion}
In this study we aim to introduce a reporting framework for reproducible and interactive research, based on technologies and methods applied to some of the recent projects in our research group (RoR). Several tools were described to publish datasets and analytical codes, all centered and managed by the R Language Software. 

The concept of our framework was initially based on some guidelines already published~\cite{Laine2007, biostatistics}. Not many reports can be found in the literature on the use of reproducible research framework in healthcare. Some researchers do publish their datasets or codes, but generally they are published separately. In our framework we tried to approach different aspects of reproducibility, rather than just the data, that is connectivity~\cite{Peng2007}, dissemination and licensing with a framework that is constituted of free and friendly technologies, facilitating \textit{replication} and improving \textit{transparency} of results.

Some of the tools we showcased here have been extensively discussed and used for the development of projects by many investigators. Github, for instance, has been extremely used by data analysts and programmers, as well as Dyrad and Figshare, given the increased amount of data being stored in clouds. However, this advances have not been observed as often in healthcare research, specifically when it comes to Big Clincal Data and replication of health researches protocols \cite{Groves}. 

Although our proposed framework is still in progress and needs to be improved, we emphasize its ability not only for sharing data and codes in a safe way, but also connecting and disseminating information through free and user friendly technologies. We believe that only by sharing and comparing methods a consensus of framework can be created. Therefore, this model proposed can help towards the standardization of reproducible research protocols in healthcare, aggregating value not only for research, but also for innovation and clinical practice.

\end{multicols}

\end{document}